\documentclass[aps,prd,preprint,superscriptaddress]{revtex4-1}
\usepackage[english]{babel}
\usepackage[utf8]{inputenc}
\usepackage[T1]{fontenc}
\usepackage{amsmath,amssymb,braket,slashed,mathrsfs}
\usepackage{pifont}
\usepackage{hyperref}
\usepackage{color,hyperref}
\usepackage{cleveref}
\usepackage{amsmath} 
\usepackage{graphicx}
\usepackage{booktabs}
\usepackage{multirow}
\usepackage{subfigure}
\usepackage{float}

\newcommand{\ba}{\begin{eqnarray}}
\newcommand{\ea}{\end{eqnarray}}
\newcommand{\be}{\begin{equation}}
\newcommand{\ee}{\end{equation}}
\newcommand{\bi}{\begin{itemize}}
\newcommand{\ei}{\end{itemize}}

\newcommand{\nn}{\nonumber}

\newcommand{\innovation}{Collaborative Innovation Center of Quantum Matter, Beijing 100871, China}
\newcommand{\chep}{Center for High Energy Physics, Peking University, Beijing 100871, China}
\newcommand{\pkuphy}{School of Physics, Peking University, Beijing 100871,
China}
\newcommand{\KeyLab}{State Key Laboratory of Nuclear Physics and Technology,
Peking University, Beijing 100871, China}
\newcommand{\Uconn}{Department of Physics, University of Connecticut, Storrs, CT 06269, USA}
\newcommand{\RBRC}{RIKEN-BNL Research Center, Brookhaven National Laboratory, Building 510, Upton, NY 11973}

\begin{document}

\title{Field sparsening for the construction of the correlation functions in lattice QCD} 

\author{Yuan Li}\affiliation{\pkuphy}
\author{Shi-Cheng Xia}\affiliation{\pkuphy}
\author{Xu Feng}\affiliation{\pkuphy}\affiliation{\innovation}\affiliation{\chep}\affiliation{\KeyLab}
\author{Lu-Chang Jin}\affiliation{\Uconn}\affiliation{\RBRC}
\author{Chuan Liu}\affiliation{\pkuphy}\affiliation{\innovation}\affiliation{\chep}

\date{\today}

\begin{abstract}
    Two field-sparsening methods, namely
    the sparse-grid method and the random field selection method, are used in
    this paper for the construction of the 2-point and 3-point correlation functions
    in lattice QCD. We argue that, due to the high correlation among the 
    lattice correlators at different field points 
    associated with source, current, and sink locations, 
    one can save a lot of computational time by 
    performing the summation over a subset of the lattice sites.
    Furthermore, with this strategy,
    one only needs to store a small fraction of the full quark propagators. It is found that 
    the number of field points can be reduced by a factor of $\sim$100 for the
    point-source operator and a factor of $\sim$1000 for the Gaussian-smeared operator, 
    while the uncertainties of the correlators only increase by $\sim$15\%.
    Therefore, with a modest cost of the computational resources, 
    one can approach the precision of the all-to-all correlators using the
    field-sparsening methods.
\end{abstract}

\maketitle

\section{introduction}

Lattice QCD provides a non-perturbative approach to the numerical 
solution of Quantum Chromodynamics (QCD), which is believed to be 
the basic theory of strong interactions among quarks and gluons. 
With the development of the cutting-edge supercomputers, the new algorithms
and the advanced methodologies, lattice QCD is now playing an increasingly important role in
the understanding of low-energy QCD. 

 A typical way to gain a better
 efficiency in a numerical lattice QCD calculation is to reduce the 
 redundant costs. Here are some examples. 
\bi
\item To guarantee an accurate generation of a quark propagator,
    the residual criterion in the conjugate gradient (CG) inversion for the quark propagator is
    usually set to $10^{-8}$ or even smaller. 
    When realizing that most of the CG iterations can be saved, the all mode
    averaging (AMA) technique~\cite{Blum:2012uh} is proposed, where the residual criterion is raised to a
    level of $\sim 10^{-4}$. In this way the computational time is significantly
    reduced while the physical quantities can still be obtained with no bias by
    performing a correction, which compensates the systematic effects from the
    approximated propagators by adding the difference between some samples of
    the precise correlators and the approximated ones.
\item  In the calculation of hadronic light-by-light contributions to the muon
    anomalous magnetic moment, a four-point hadronic function with vector-currents is required. 
    To construct such a four-current correlator, one expects a spacetime summation over
    the locations of at least three of the four currents which is very challenging numerically. 
    Realizing that in the connected diagram, when the locations of two vector currents 
    are separated with a spacetime distance $r$, 
    the hadronic function falls exponentially with the increase of
    $r$, an importance sampling is introduced to evaluate the stochastic sum over
    $r$ efficiently~\cite{Blum:2015gfa}. Therefore, in the important regions where $r\lesssim1$ fm, 
    the summation is run in a complete way while in the other regions $r>1$ fm,
     the contributions are calculated with a probability of
    $p(r)\propto 1/r^{3.5}$. In this way, much less computational resources are
    spent in the very long distance region, where the lattice correlation functions 
    mainly contribute noise rather than the signal.
\item In many cases, it is appealing to utilize the translational invariance and
    construct the correlation function using
    the all-to-all propagators~\cite{Foley:2005ac}. As a result, the information over the whole
    spacetime volume is summed and one expects to gain a good precision for the
    correlation function. On the other hand, generating all-to-all propagators is 
    quite time consuming. Since the correlation functions from the same
    configuration is highly correlated, one can achieve nearly the same  
    precision by averaging over part of the source and sink locations.
    Such kind of techniques, called {\em field-sparsening methods} here, 
    are the main focus of this paper. 
\ei

The field-sparsening techniques have been studied by Detmold and Murphy in Ref.~\cite{Detmold:2019fbk}, 
where a sparse-grid technique is introduced. 
An earlier application of the sparse grid can be traced back to
a study from $\chi$QCD collaboration in 2010~\cite{Li:2010pw}.
In this work, in addition to the sparse-grid approach, we developed another
field-sparsening approach, which we will refer to as the random field selection method.
We will study and compare these two methods in some detail.
The paper is organized as follows. We start with
Sect.~\ref{sect:field_selection} by introducing
the two field-sparsening techniques mentioned above. In
Sect.~\ref{sect:2point_function} we discuss the
results of the 
2-point functions for pion and proton with both point-source and
Gaussian-smeared-source operators. The advantages and disadvantages for each
field-sparsening method are also discussed. In Sect.~\ref{sect:model} we employ a simple model to analyze
the sources of the uncertainties for the random field selection method. 
In Sect.~\ref{sect:3point_function} we extend the study to the 3-point function, where the
proton axial charge $g_A$ is used as an example to demonstrate the efficiency of
the field sparsening methods. 

\section{Field sparsening methods}
\label{sect:field_selection}

In the lattice QCD calculation of a generic $n$-point function,
\be
\sum_{\vec{x}_1,\vec{x}_2,\cdots,\vec{x}_{n-1}\in
\Lambda_{\mathrm{full}}}\langle O_1(x_1)O_2(x_2)\cdots O_n(x_n)\rangle
\;,
\ee
with $O_i(x_i)$ being the interpolating operators at temporal-spatial 
point $x_i=(x^0_i,\vec{x}_i)$ and $\Lambda_{\mathrm{full}}$ the full set
of spatial lattice points,
one needs to perform the volume summation $(n-1)$ times. 
This results in a computational cost of $(L^3)^{n-1}$ in the quark contraction, 
 with $L$ being the spatial lattice size. 
 If one wants to gain a better precision by making another spatial-volume average over the
 locations of $x_n$, then the cost becomes $(L^3)^{n}$. The typical size of $L^3$ is about
 $10^4$-$10^6$ for practical lattice QCD simulations. Given a relatively large lattice, 
 the complexity of $O(L^6)$ usually exceeds the capability of the current lattice QCD calculations.
 In many cases, the techniques of all-to-all propagators~\cite{Foley:2005ac} or sequential-source
 propagators~\cite{Martinelli:1988rr} are used to reduce the computational costs. 
 On the other hand, there are some limitation for the usage of these propagators. 
 For example, the all-to-all propagators work more efficiently in the mesonic sector 
 than in the baryonic sector. For the sequential-source propagators, 
 although it allows to reduce a complexity of $O(L^6)$ to a level of $O(L^3)$,
 the computational cost can increase dramatically if one wants to build the sequential-source propagators
 with multiple time slices,  momentum insertions and gamma matrix structures.

 Given a gauge configuration from Monte Carlo simulation, 
 the correlation functions at different source and sink
 locations are usually highly correlated statistically. 
 Therefore one can save computational time by only summing over a small subset of all the possible
 the source or sink locations. In fact, as will be shown below, 
 utilizing less source locations one can efficiently reduce
 the costs to generate the quark propagators. For each propagator, one can
 use less sink locations and reduce the computational cost for the Wick
 contractions in the construction of the correlation functions. 
 Since the numbers of both source and sink locations are reduced, 
 it also saves the disk space to store these quark propagators 
 and also reduces the pressure for the input and output (I/O)
 of the large-size data on the supercomputers. It is the task of this paper to
 show how much fewer location points one can use to maintain a
 comparable precision of the correlation functions that use the full location points.

For simplicity, we start with the 2-point correlation function as an example to
introduce the field-sparsening techniques. The standard 2-point correlation 
function with zero spatial momentum insertion is written as
\be
\label{eq:2-point}
C_{\mathrm{full}}(t)=\sum_{\vec{x}\in\Lambda_{\mathrm{full}}}\langle
O(\vec{x},t_0+t)O^\dagger(\vec{x}_0,t_0)\rangle,
\ee
where the subscript ``full'' indicates that the summation of $\vec{x}$ runs over
all the sink location points.
By using field sparsening, one can replace the summation $\sum_{\vec{x} \in
\Lambda_{\mathrm{full}}}$ by $\frac{L^3}{N_\Lambda}\sum_{\vec{x}\in\Lambda}$,
where $\Lambda$ is a subset of $\Lambda_{\mathrm{full}}$, which contains only
$N_\Lambda$ location points. Due to the high correlation in the lattice data,
we expect that the replacement does not increase the noise much but reduces the
propagator storage and contraction time for modest size $N_\Lambda$. 
In Eq.~(\ref{eq:2-point}) only one source location $(\vec{x}_0,t_0)$ is used. In
principle one can use multiple source locations and write the correlation
functions as
\be
\label{eq:2-point1}
C_{\mathrm{sparse}}(t)=\frac{L^3}{N_\Lambda
N_{\Lambda_0}N_{\Lambda_t}}\sum_{\vec{x}\in\Lambda}\sum_{\vec{x}_0\in\Lambda_0}\sum_{t_0\in\Lambda_t}\langle
O(\vec{x},t_0+t)O^\dagger(\vec{x}_0,t_0)\rangle,
\ee
where the source spatial location takes the value from the set $\Lambda_0$ and
source time slice takes the value from $\Lambda_t$. The size of the set
$\Lambda_0$ and $\Lambda_t$ is given by $N_{\Lambda_0}$ and $N_{\Lambda_t}$,
respectively.
In this work we will compare two
different sets for $\Lambda$ ($\Lambda_0$), namely the sparse-grid method
and the random field selection method, and determine the optimal values
for $N_\Lambda$ and $N_{\Lambda_0}$ in each case.

\subsection{Sparse-grid method}
Following
the sparse-grid method introduced in Ref.~\cite{Detmold:2019fbk}, 
the set $\Lambda$ is chosen as,
\be
\mbox{type I: }\{(n_1,n_2,n_3)\big|0\le n_i<L;\,\, n_i =0\ (\operatorname{mod}
k) \},
\ee
where $k$ is an integer factor of $L$. By using this setup, a $L^3$-point
spatial lattice is reduced to a $(L/k)^3$-point one. 
For all the time slices $t_0$ and $t_0+t$ used in Eq.~(\ref{eq:2-point}), 
the same sparse grid set is implemented.
One can also extend the type of the sparse grid to
\begin{equation}
\begin{split}
    \mbox{type II : }&\{(n_1,n_2,n_3)\big|0\le n_i<L;\,\, n_i
    =0(\operatorname{mod}k); \,\, n_1+n_2+n_3= 0(\operatorname{mod}2k)\},\\
    \mbox{type III : }& \{(n_1,n_2,n_3)\big|0\le n_i<L;\,\, n_i
    =0(\operatorname{mod}k); \,\, n_i+n_j=0(\operatorname{mod}2k),i\neq j  \}.
\end{split} 
\end{equation}
With the above definitions, we have
$N_\Lambda=\frac{L^3}{k^3},\frac{L^3}{2k^3},\frac{L^3}{4k^3}$ for type I, II and
III, respectively.
In our numerical study, we use a lattice gauge ensemble with $L=24$ and pick up
15 values for $N_\Lambda$. For convenience, these 15 cases are labelled by
an integer denoted as $N_{\mathrm{th}}$, running from $0$ to $14$, and the
corresponding values for $N_\Lambda$ are given by the following list,
\be
\label{eq:choice_N_Lambda}
N_{\Lambda}(N_{\mathrm{th}})=\left\{24^3,12^3,\frac{12^3}{2},\frac{12^3}{4},6^3,\frac{6^3}{2},4^3,\frac{6^3}{4},\frac{4^3}{2},3^3,\frac{4^3}{4},2^3,4,2,1
\right\}.
\ee
It means that we have $N_\Lambda=24^3$ for $N_{\mathrm{th}}=0$ and $N_\Lambda=1$
for $N_{\mathrm{th}}=14$.

Note that the sets of type I, II, III always include the location point of
$(n_1,n_2,n_3)=(0,0,0)$. We therefore can consider it as a reference point. 
To reduce the correlation in the lattice calculation,
for each configuration, one can shift the reference point randomly with
$L^3/N_{\Lambda}$ choices.

\subsection{Random field selection}
Another choice of $\Lambda$ is called random field selection, with $\Lambda$
chosen as
\be
\Lambda=\{(n_1,n_2,n_3)\big|0\le n_i<L;\,\, \mbox{$n_i$ are random numbers} \}.
\ee
The random numbers for $n_i$ vary when the time slice $t$ and configuration
trajectory change. In principle, we can use any value for $N_\Lambda$. To favor
a comparison with the sparse-grid method, we use the same choices of $N_\Lambda$
as that in Eq.~(\ref{eq:choice_N_Lambda}).

\section{2-point function}
\label{sect:2point_function}

In this section we will present the results for 2-point correlation functions.
The calculation is performed using a gauge ensemble with
$2+1+1$ clover-improved Wilson twisted mass quarks, generated by the ETM
Collaboration~\cite{Alexandrou:2018egz}. 
The lattice volume is $24^3\times48$ with a pion mass $m_\pi\approx350$ MeV and a lattice spacing $a\approx0.093$ fm. In total 91 configurations are utilized in this analysis.

\begin{figure}[h]
	\centering
		\subfigure{\includegraphics[width=0.45\textwidth]{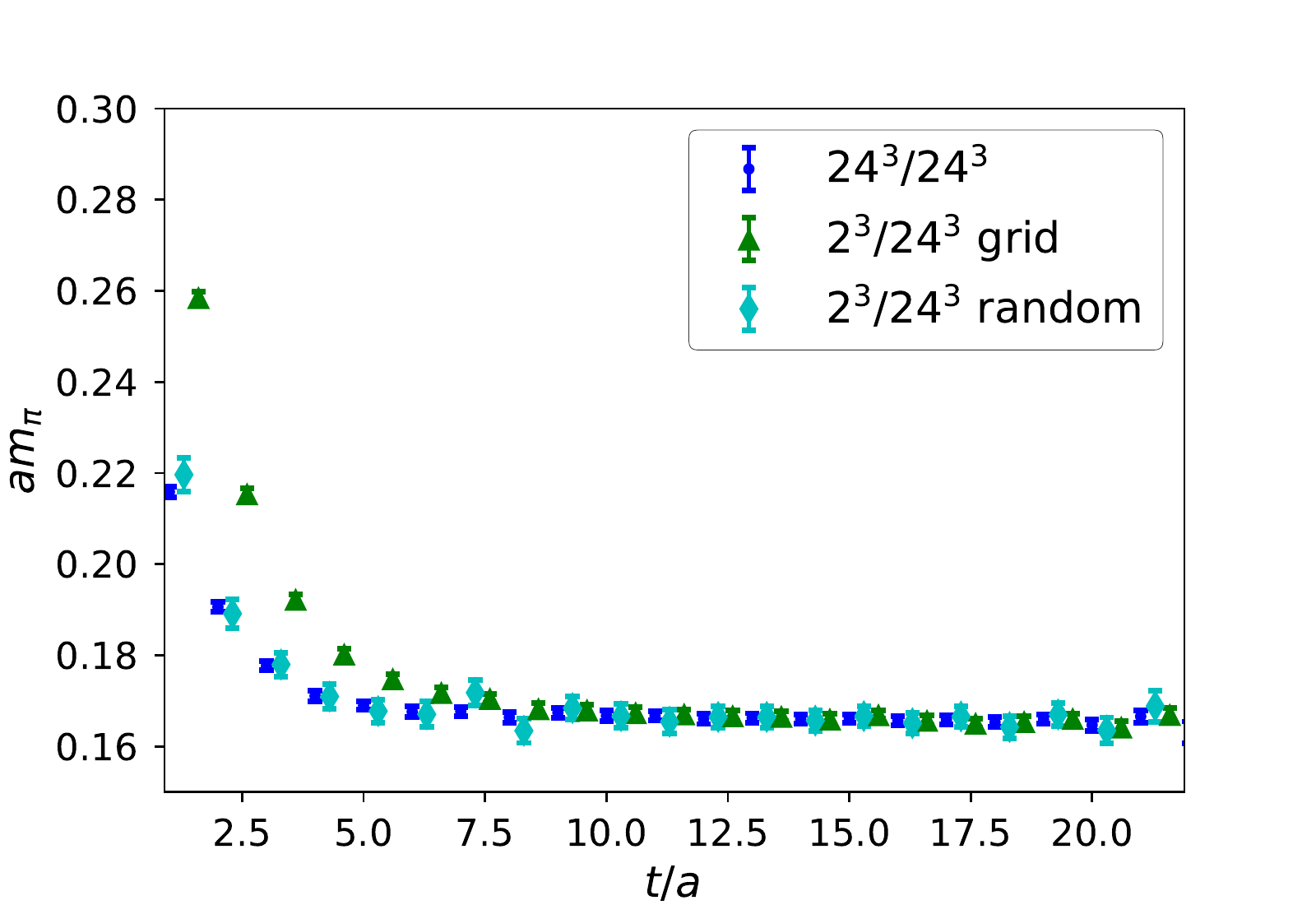}}\hspace{5mm}
		\subfigure{\includegraphics[width=0.45\textwidth]{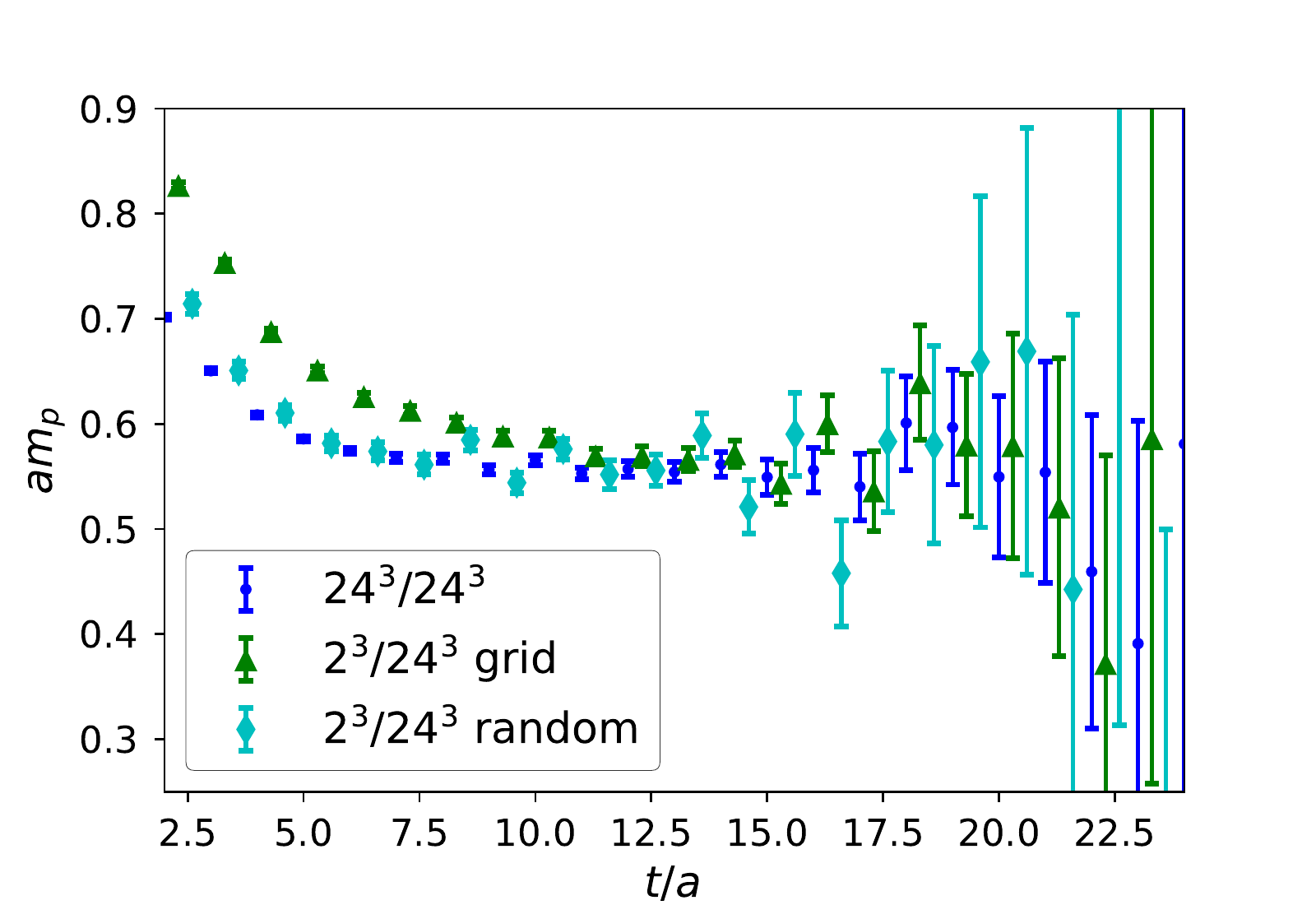}}\hspace{5mm}
    \caption{Effective masses for the pion (left panel) and the proton (right
    panel) with full set $\Lambda_\mathrm{full}$ (blue), sparse-grid (dark
    green) and random-field selection (cyan). 
    For the two field-sparsening methods, we use $N_{\Lambda}=2^3$. The green and
    cyan data points are shifted horizontally for an easier comparison.}
    \label{fig:effective_mass}
\end{figure}

We use Gaussian-smeared-source
propagators to construct the smeared-source smeared-sink 2-point functions
\ba
C_\alpha(t)=\frac{L^3}{N_\Lambda
N_{\Lambda_0}N_{\Lambda_t}}\sum_{\vec{x}\in\Lambda}\sum_{\vec{x}_0\in\Lambda_0}\sum_{t_0\in\Lambda_t}\langle
O_\alpha(\vec{x},t_0+t) O_\alpha^\dagger(\vec{x}_0,t_0)\rangle
\ea
with $\alpha=\pi$ for pion and $\alpha=p$ for proton. 
We use 24 time slices for $t_0$. For each time slice, we use 8 random source
locations for $\vec{x}_0$. Thus we have $N_{\Lambda_t}=24$ and
$N_{\Lambda_0}=8$. For the sink location $\vec{x}$, it can be summed
over the full set $\Lambda_{\mathrm{full}}$ or field-sparsening set $\Lambda$.
Fig.~\ref{fig:effective_mass} shows the effective masses for the pion (left panel) and the proton (right
panel) with the point set $\Lambda_{\mathrm{full}}$ and $\Lambda$ at
$N_\Lambda=2^3$. The effective masses $m_\alpha$ at the time slice $t$ are obtained using
$C_\alpha(t)$ and $C_\alpha(t+1)$ as inputs. For the two field-sparsening
methods, the data points are slightly shifted horizontally to favor a 
comparison with that from the full set.

From Fig.~\ref{fig:effective_mass} a clear enhancement of the excited-state
contamination is found in the sparse-grid method at $N_{\Lambda}=2^3$. This is due to the mixing of the
hadron states with high momenta. Let us take the sparse-grid set type I as an
example. The summation over $\Lambda$ can be written as
\be
\frac{L^3}{N_{\Lambda}}\sum_{\vec{x}\in\Lambda}=\sum_{\vec{m}\in \Gamma}\sum_{\vec{x}\in\Lambda_{\mathrm{full}}}
e^{i\frac{2\pi}{k}\vec{m}\cdot\vec{x}},
\ee
with
$\Gamma=\{(m_1,m_2,m_3)\big|m_i =0,1,\cdots,k-1 \}$. Therefore the
higher-momentum modes with $\vec{m}\neq\vec{0}$ mix with the zero-momentum mode.
As a consequence, the excited-state contaminations 
increase as $N_{\Lambda}$ decreases. The situation becomes even more
problematic when one targets on the calculation of the correlation functions with
large momentum transfer. In this case it is possible that a state with an assigned
momentum mixes with
the states carrying smaller momenta and consequently the correlation functions are distorted
by the low-lying states.

For the random-field selection method, there is no enhancement of the excited-state
contribution. On the other hand, the statistical errors become larger due to the
random noise arising from the field sparsening. 
Note that the correlation among $C_\alpha(t)$ at various $t$ is
weakened by the random-field selection. As a result, the uncertainty for
the effective mass can be further reduced if one performs a fit over a temporal window.

 \begin{figure}[h]
	\centering
	\subfigure{\includegraphics[width=0.45\textwidth]{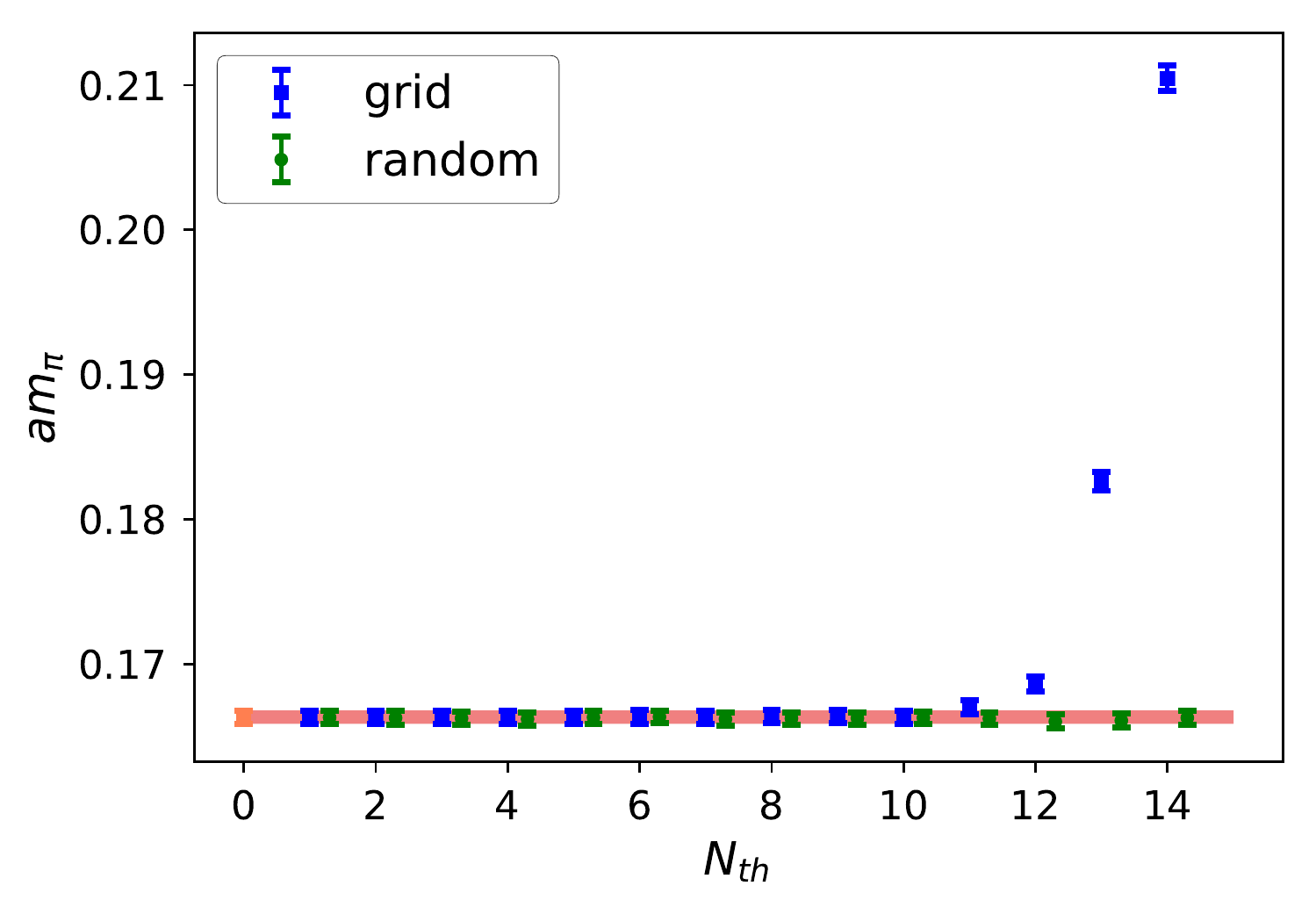}}\hspace{5mm}
	\subfigure{\includegraphics[width=0.45\textwidth]{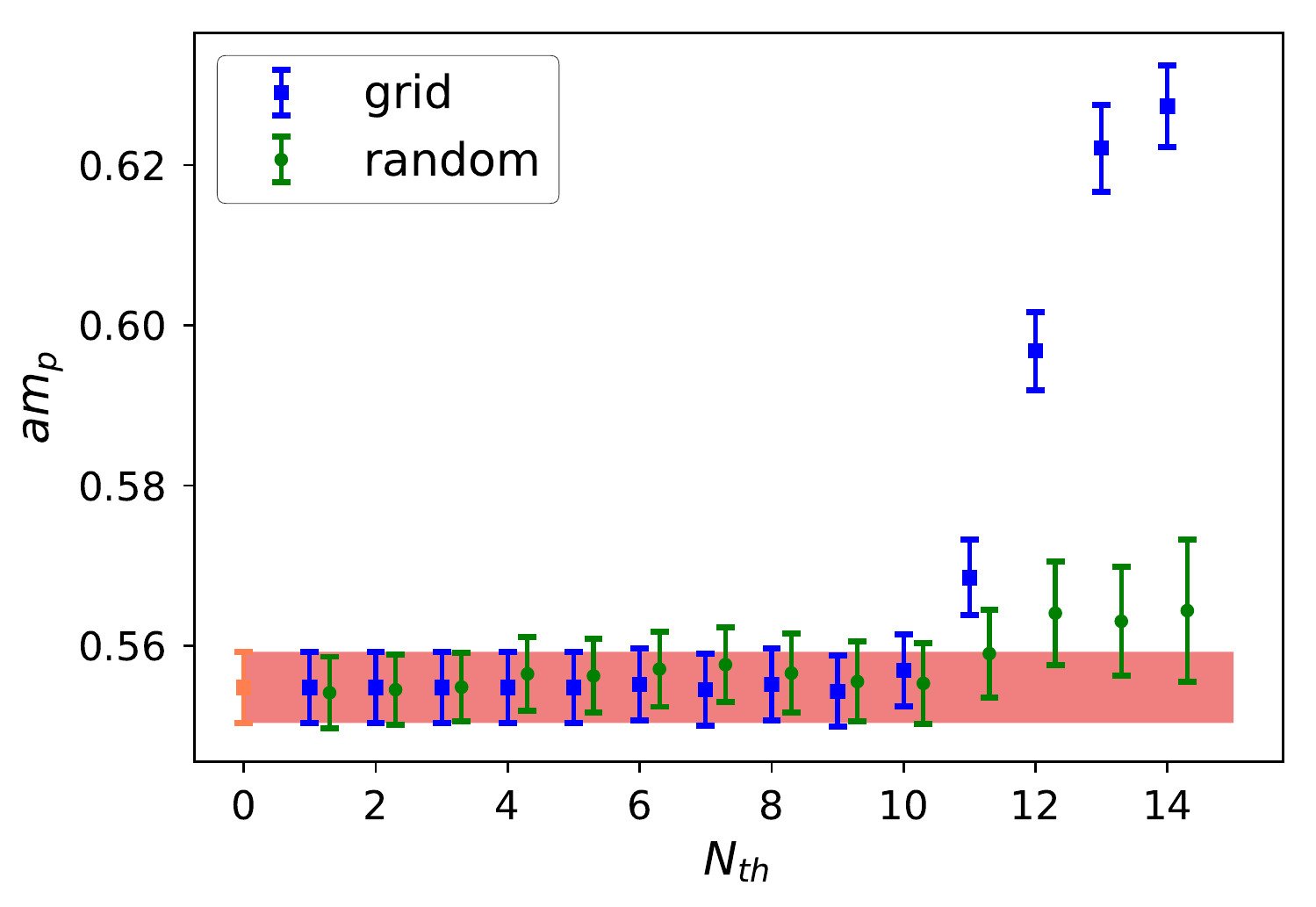}}\hspace{5mm}
     \caption{Effective masses for the pion (left panel) and proton (right panel)
     from the correlated fit as a function of $N_{\mathrm{th}}$. For each
     $N_{\mathrm{th}}$, the number of
     field points is given in Eq.~(\ref{eq:choice_N_Lambda}). The data points from
     sparse-grid method and random field selection are printed in blue and green
     color, respectively.}
	\label{fig:different_selection}
\end{figure}

 In Fig.~\ref{fig:different_selection}, we show the effective mass from the correlated fit as a function of $N_{\mathrm{th}}$, or equivalently the different choices of $N_{\Lambda}$. 
 The fitting windows are determined using the data with $\Lambda_{\mathrm{full}}$. 
 To be specific, we use the fitting window $8\le t\le 22$ for 
 the pion correlator and obtain $m_\pi=0.16632(46)$ from a correlated fit 
 with $\chi^2/\mathrm{dof}=0.96$. We use the fitting window $10\le t\le 22$ for the proton 
 and obtain $m_p=0.5548(44)$ with $\chi^2/\mathrm{dof}=0.35$. 
 We denote these effective masses as $m_\pi^{\mathrm{full}}$ and $m_p^{\mathrm{full}}$.
 The reasonable values of $\chi^2/\mathrm{dof}$ suggest that the
 excited-state contributions are well under control. 
 We thus fix these fitting windows for the field-selection cases. 
 For the sparse-grid method, the additional excited-state contributions from higher momenta 
 start to make an obvious impact on the effective mass 
 when $N_{\mathrm{th}}\ge11$ (or $N_{\Lambda}\le8$). For the random
 field selection, although the effective masses at each time slice carries larger
 errors, the uncertainties are very close to the ones from sparse-grid method after a
 correlated fit. We find that the effective masses at $N_{\mathrm{th}}=10$ ($N_{\Lambda}=16$) 
 is given by $m_\pi=0.16627(44)$ and $m_p=0.5553(51)$. 
 By using the random field selection method, one can reduce in the summation the number 
 of the sink points by a factor of $864$ ($N_{\Lambda}=L^3\to16$),  while the lattice results 
 are still very precise. For the pion effective mass, 
 the statistical uncertainty is consistent with that from $\Lambda_{\mathrm{full}}$, 
 and for the proton, the uncertainty only increases by about $16$\%.

 \begin{figure}[h]
	\centering
	\subfigure{\includegraphics[width=0.45\textwidth]{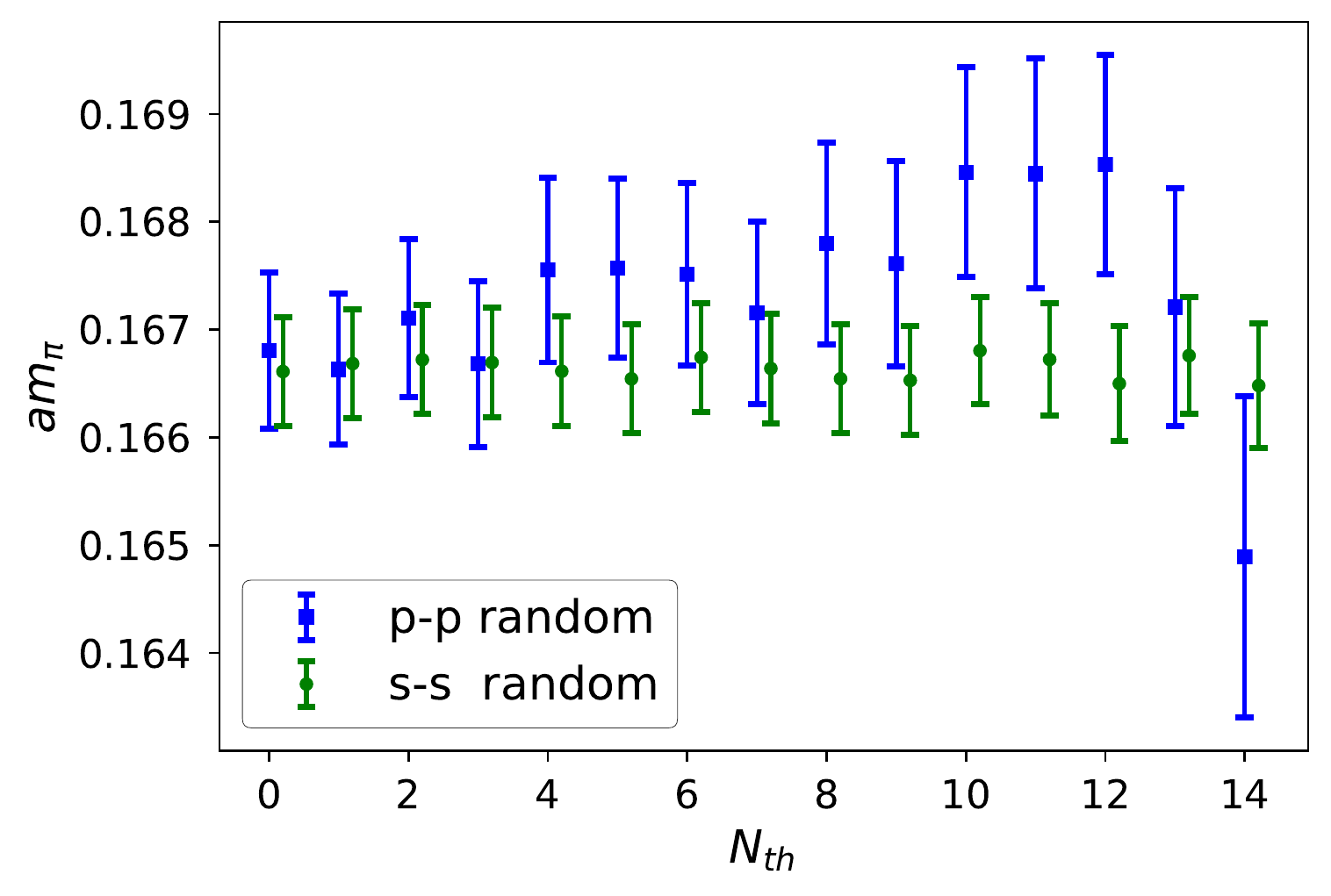}}\hspace{5mm}
	\subfigure{\includegraphics[width=0.45\textwidth]{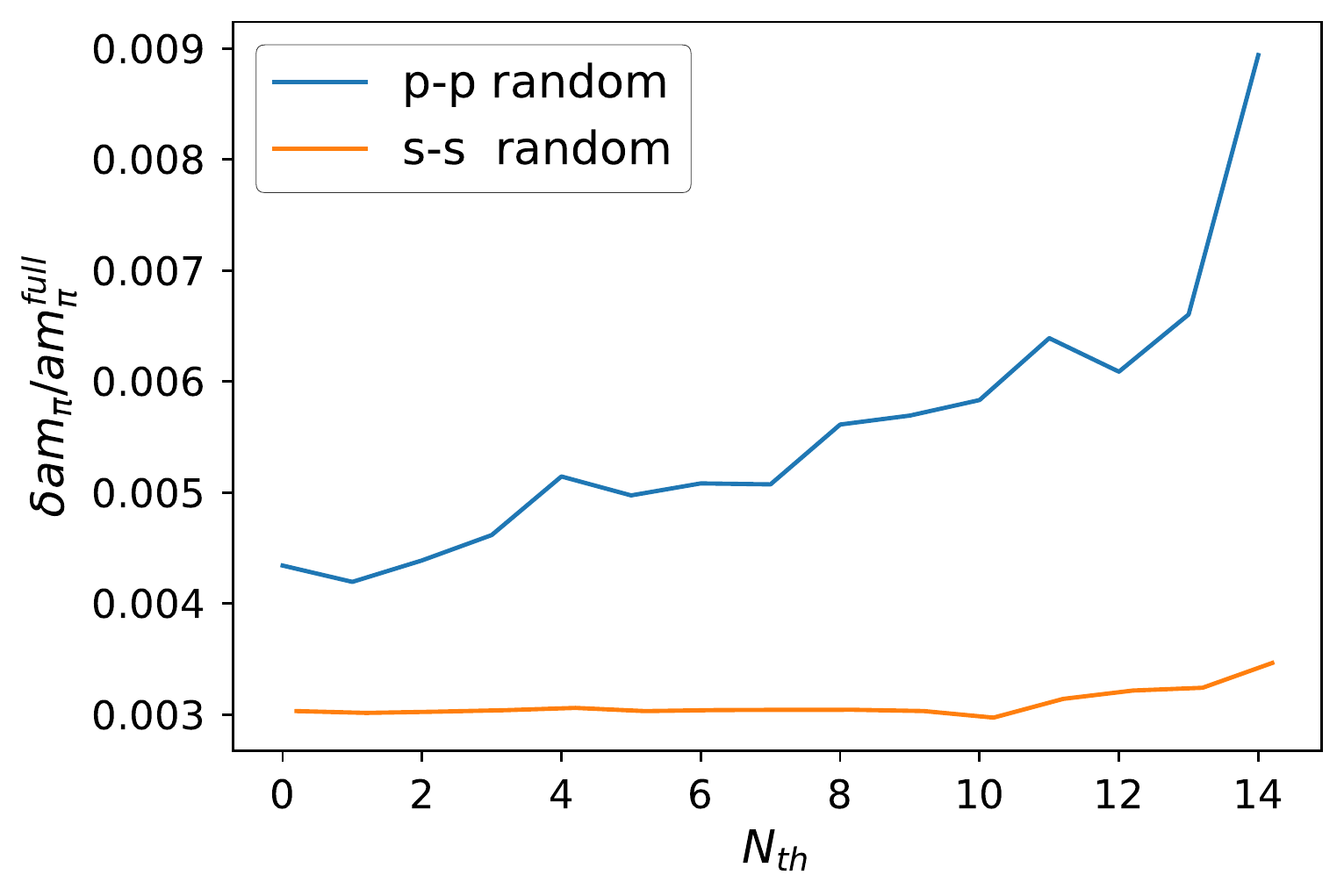}}\hspace{5mm}
	\subfigure{\includegraphics[width=0.45\textwidth]{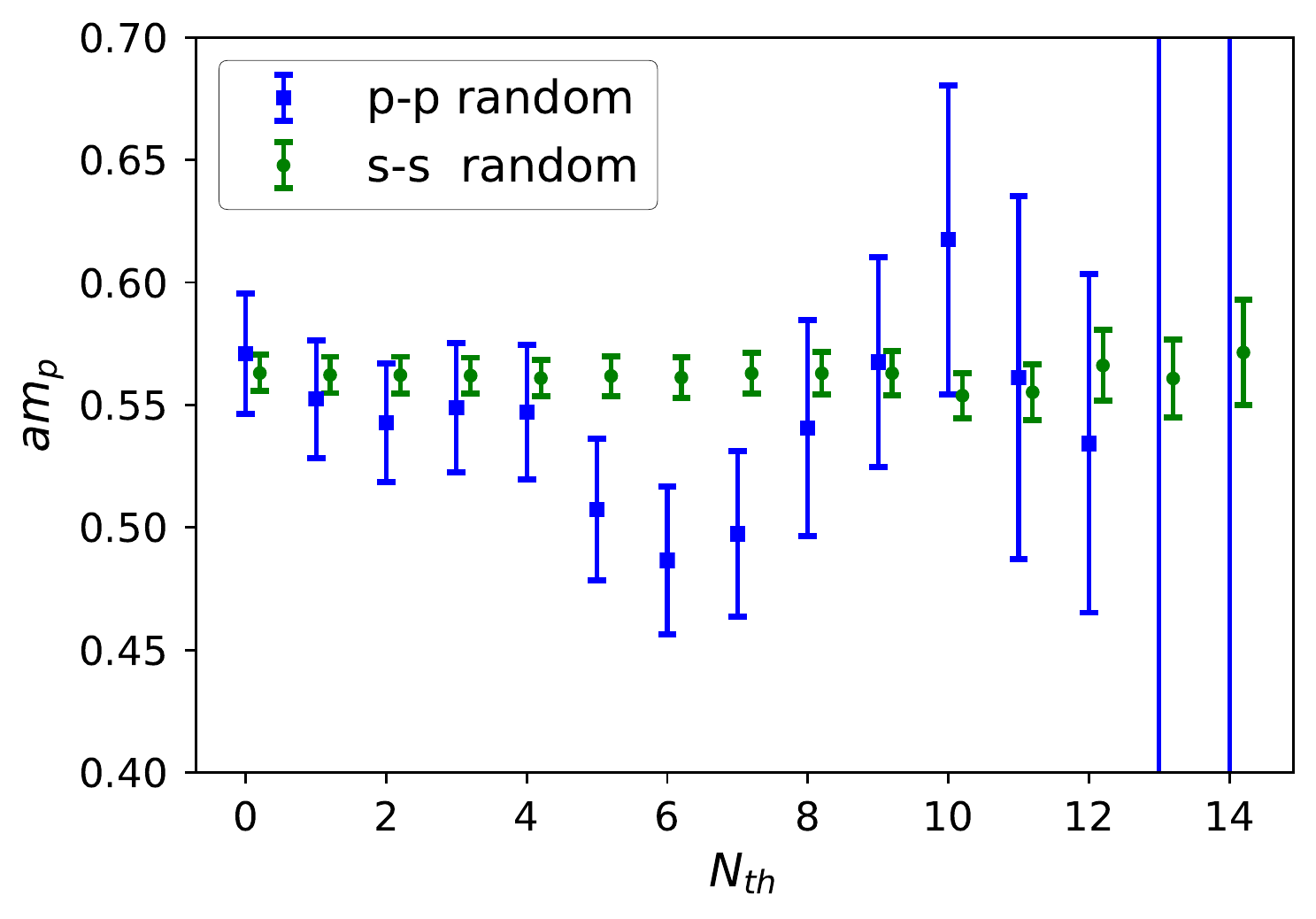}}\hspace{5mm}
	\subfigure{\includegraphics[width=0.45\textwidth]{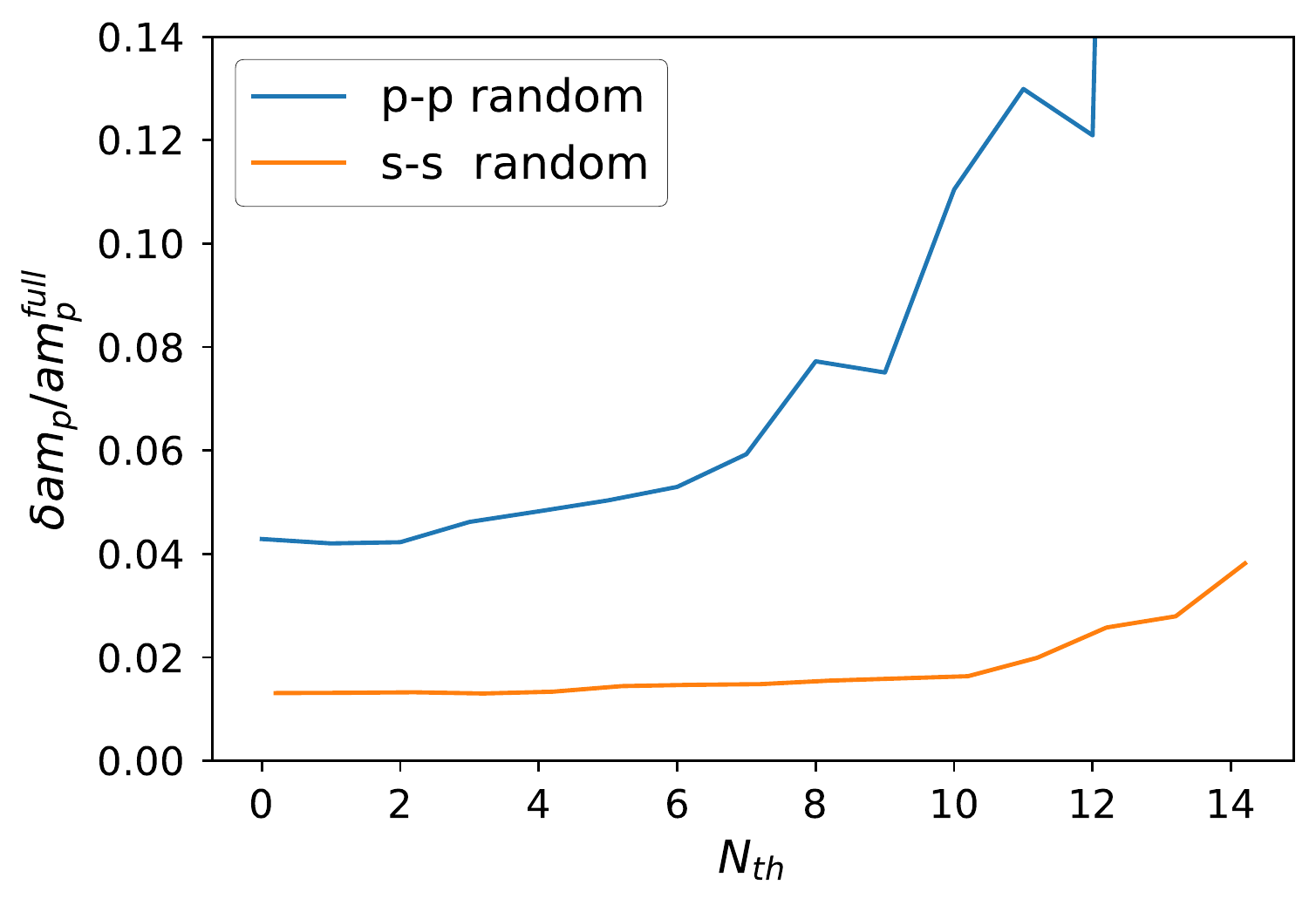}}\hspace{5mm}
     \caption{Effective masses for the pion and proton from the 
     smeared-source smeared-sink (s-s) and point-source point-sink (p-p)
     correlation functions. In the left panel, the effective masses as a function
     of $N_{\mathrm{th}}$ are shown. Here we use the random field selection. 
     In the right panel, the ratio between the statistical
     uncertainty and the effective mass $m_\alpha^{\mathrm{full}}$ for $\alpha=\pi$
     and $p$ is shown.}
	\label{fig:pip_ppVSss}
\end{figure}

The efficiency of the field sparsening likely depends on the type of interpolating
operators used in the lattice calculation. As the Gaussian-smeared-source
propagator is more correlated than the point-source propagator, we expect that the
field sparsening works more efficiently for the former case. To confirm this conjecture we calculate
the effective masses for the pion and proton using both Gaussian-smeared-source
and point-source propagators. To save the cost, we keep $N_{\Lambda_t}$ at 24
while reducing $N_{\Lambda_0}$ to 1 for both smeared-source smeared-sink (s-s) and
point-source point-sink (p-p) correlation
functions. The effective
masses from the correlated fit are shown in the left panels of Fig.~\ref{fig:pip_ppVSss}, 
 while the ratios between the statistical uncertainty and the effective 
 masses $m_\alpha^{\mathrm{full}}$ for $\alpha=\pi$
 and $p$ are shown in the corresponding right panels. With the same statistics,
the results of p-p correlators are noisier than that of the s-s ones. The field-sparsening
method works less efficiently in the p-p corelators as we expect. 
 Nevertheless, we find that at $N_{\mathrm{th}}=5$ ($N_\Lambda=108$)
 the uncertainty of the p-p effective mass increases only 15\% for the pion and
 18\% for the proton. This implies that one can reduce the field points by a factor
 of 128 in trade with a small increase of the statistical error. Two order of
 magnitude reduction in the propagator storage and I/O data transfer
 would make a typical lattice calculation much easier.

\section{Noise from random field selection method}
\label{sect:model}

 When using the random field selection method, the correlation function receives two
 types of the noise. The first is the gauge noise, $\delta_{\mathrm{gauge}}$, 
 and the second is the noise from the selection of the random field points, $\delta_{\mathrm{rand}}$. 
 The increase of the gauge noise is an inevitable
 price when the number of the field points is reduced in the summation.
 Since the lattice data at the different points are highly correlated, we expect the gauge noise only
 increases mildly. We therefore focus on the estimation of $\delta_{\mathrm{rand}}$.

 We write the time dependence of the 2-point correlation function as
\ba
\label{eq:2-point_func}
C(\vec{x},t)&=&\langle O(\vec{x},t)O^\dagger(\vec{0},0)\rangle
\nn\\
&\doteq&\frac{1}{L^3}\sum_{\vec{p}}e^{i\vec{p}\cdot\vec{x}}
\frac{\tilde{A}(\vec{p})}{2E}\left(e^{-Et}+e^{-E(T-t)}\right),
\ea
where the sign of $\doteq$ reminds us that the excited-state contributions are
neglected at sufficiently large $t$. The energy $E$ satisfies the dispersion
relation $E=\sqrt{m^2+\vec{p}^2}$ with $m$ the hadron's mass.

To fully understand the impact on the uncertainty of the correlation function from the
random field selection,
one needs to determine the weight function $\tilde{A}(\vec{p})$. Considering the
fact that we have used the Gaussian-smeared source and sink in our calculation, we
assume that the weight function in the coordinate space is given by a Gaussian
distribution
\be
	A(\vec{x})= A_0 e^{-\vec{x}^2/\sigma^2}.
\ee
Then the weight function in the momentum space is given by the Fourier
transformation of $A(\vec{x})$
\be
\label{eq:model}
\tilde{A}(\vec{p})=\int d^3\vec{x}\,
e^{-i\vec{p}\cdot\vec{x}}A(\vec{x})=A_0\left(\frac{\pi}{\sigma}\right)^{\frac{3}{2}}e^{-\frac{\sigma^2}{4}\vec{p}^2}.
\ee

 \begin{figure}[h]
	\centering
	\includegraphics[width=0.45\textwidth]{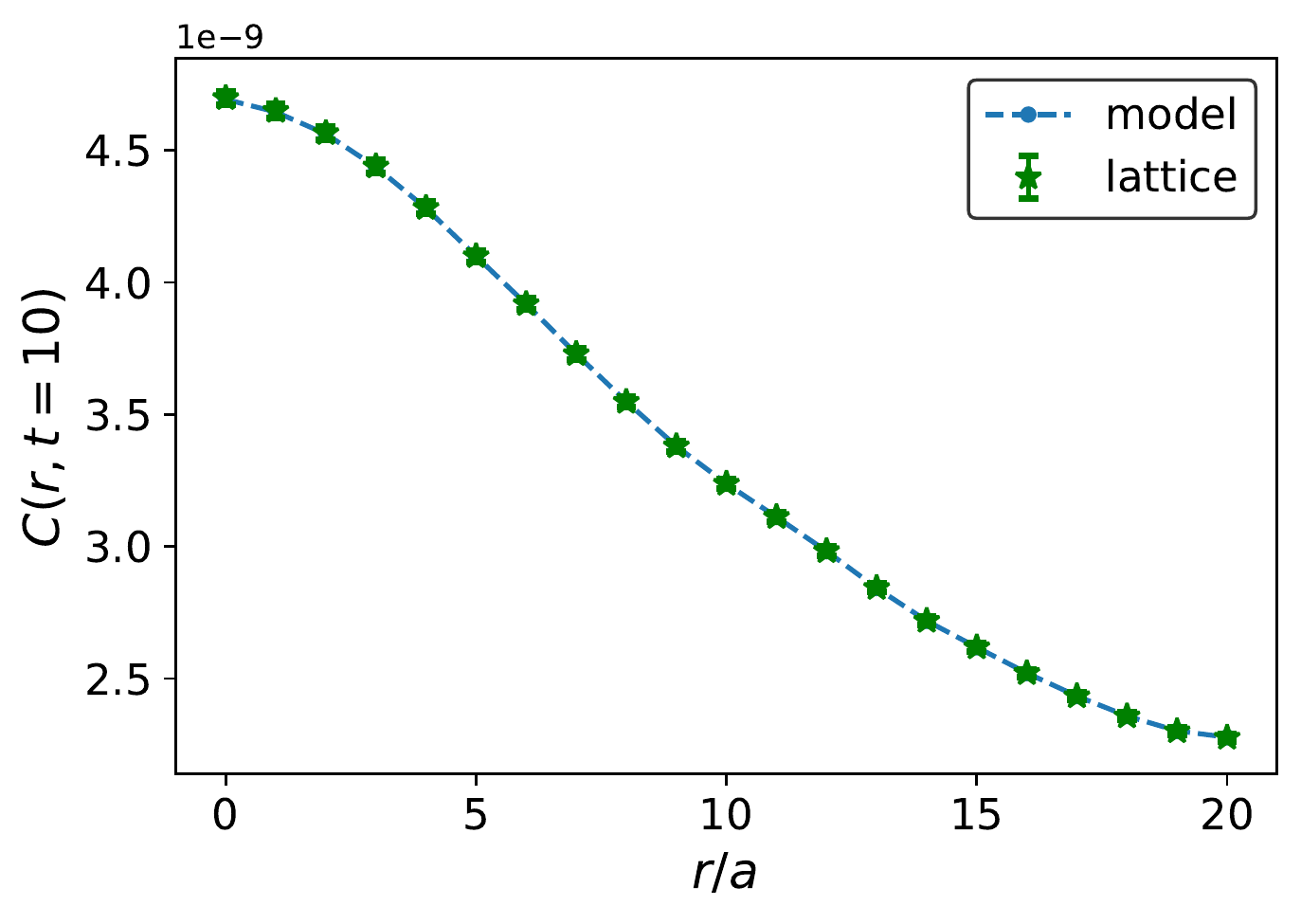}
	\caption{Lattice results of $C(r,t)$ at $t=10$ together with a fit
     to Eq.~(\ref{eq:2-point_func}).}
	\label{fig:fit_position}
\end{figure}

In the following context we use the pion correlation function as an example.
The analysis for the case of the proton is similar.
Plugging Eq.~(\ref{eq:model}) into Eq.~(\ref{eq:2-point_func}) we can use the resulting
formula to fit the pion lattice correlator with three free parameters, namely $A_0$, $\sigma$ and $m$. 
 We obtain $m=0.165(1)$ which is consistent with the effective 
 mass $m_\pi^{\mathrm{full}}$ calculated before.
 Furthermore, we can obtain the result for $C(r,t)$ by averaging 
 the lattice data of $C(\vec{x},t)$ in a range of $r\le|\vec{x}|<r+1$.
 In Fig.~\ref{fig:fit_position} we show the $C(r,t)$ at
 $t=10$ together with the fitting curve. The good agreement suggests that the functional
 form given in Eq.~(\ref{eq:2-point_func}) describes the lattice data well at large distances.

\begin{figure}
	\includegraphics[width=1\textwidth]{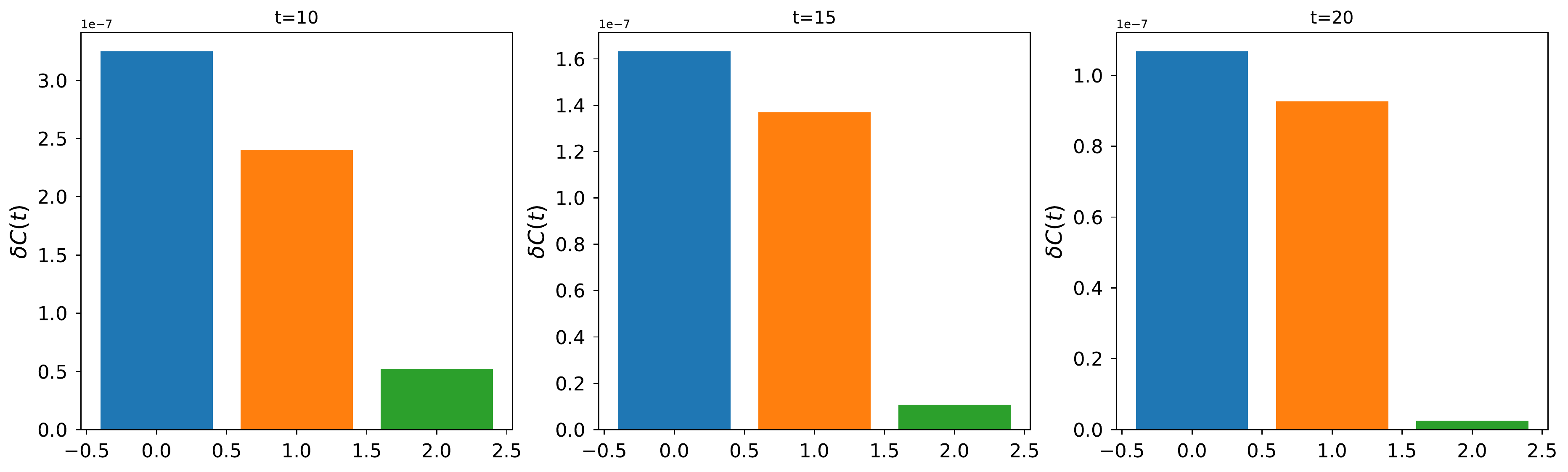}
	\caption{
    Uncertainties for the three types of 2-point functions at $t=10$, $15$ and $20$. The blue bar indicates the
    uncertainty of the correlation function with random field selection for
    $N_{\Lambda}=1$. The orange bar shows the uncertainty for
    the data using $\Lambda_{\mathrm{full}}$. 
    The green bar shows the $\delta_{\mathrm{rand}}$ only.}
	\label{fig:err_isolation}
\end{figure}

As a next step, we use the parameters $A_0$, $\sigma$ and $m$ and
Eqs.~(\ref{eq:2-point_func}) and (\ref{eq:model}) to
construct the correlation function $\hat{C}(\vec{x},t)$. As the gauge noise is
eliminated in $\hat{C}(\vec{x},t)$, we can isolate the noise
$\delta_{\mathrm{rand}}$ by replacing the lattice correlator $C(\vec{x},t)$ with $\hat{C}(\vec{x},t)$
in the random field selection method.
In Fig.~\ref{fig:err_isolation} we compare the uncertainties for the three different types of
2-point functions at $t=10$, $15$ and $20$: The blue bar indicates the
uncertainty of the correlation function with random field selection method with
$N_{\Lambda}=1$. The orange bar shows the uncertainty for the data 
with $N_\Lambda=\Lambda_{\mathrm{full}}$ while
the green bar indicates the size of $\delta_{\mathrm{rand}}$ only. 
It is noticed that the size of $\delta_{\mathrm{rand}}$ is much smaller 
 than the gauge noise and thus can be safely neglected.
 We have thus reached a conclusion that the precision of the pion 2-point functions 
 using the random field selection method is equally good as that using the sparse-grid method. 
 The random field selection is theoretically cleaner as there 
 is no enhancement of the excited-state contamination.

\section{3-point function}
\label{sect:3point_function}

In this section, we extend the random field selection method to the 3-point function,
where we calculate the proton axial charge $g_A$ as an example which is one of
the most fundamental quantities in nuclear physics. A global
average of the lattice results of $g_A$ can be found in the latest FLAG review~\cite{Aoki:2019cca}.
In this study we mainly focus on the efficiency of the random field selection method rather than the
 full control of various systematic effects for $g_A$. 
 We start with the 3-point and 2-point functions as
\begin{equation}
    C^{\mathrm{3pt}}(t_i,t_s)=\frac{L^6}{N_{\Lambda} N_{\Lambda_s}
    N_{\Lambda_0}N_{\Lambda_t}}\sum_{\vec{x} \in
    \Lambda}\sum_{\vec{x}_s\in\Lambda_s}\sum_{\vec{x}_0 \in
    \Lambda_0}\sum_{t_0\in\Lambda_t}\langle \mathcal{P}[O_p(t_0+t_s,\vec{x}_s) A_3(t_0+t_i,\vec{x})
    O_p^{\dagger}(t_0,\vec{x}_0)]\rangle
\label{threep}
\end{equation}
and
\begin{equation}
    C^{\mathrm{2pt}}(t_s)=\frac{L^3}{N_{\Lambda_s}
    N_{\Lambda_0}N_{\Lambda_t}}\sum_{\vec{x}_s \in
    \Lambda_s}\sum_{\vec{x}_0 \in \Lambda_0}\sum_{t_0\in\Lambda_t}\langle
    \mathcal{P}\left[O_p(t_s+t_0,\vec{x}_s)
    O_p^{\dagger}(t_0,\vec{x}_0)\right]\rangle,
\label{twop}
\end{equation}
where $O_p$ is the Gaussian-smeared operator for the proton, $A_3$ is the axial vector current with the polarization in the $z$
direction and $\mathcal{P}$ is the spin projection operator.  

 \begin{figure}
 	\includegraphics[width=0.45\textwidth]{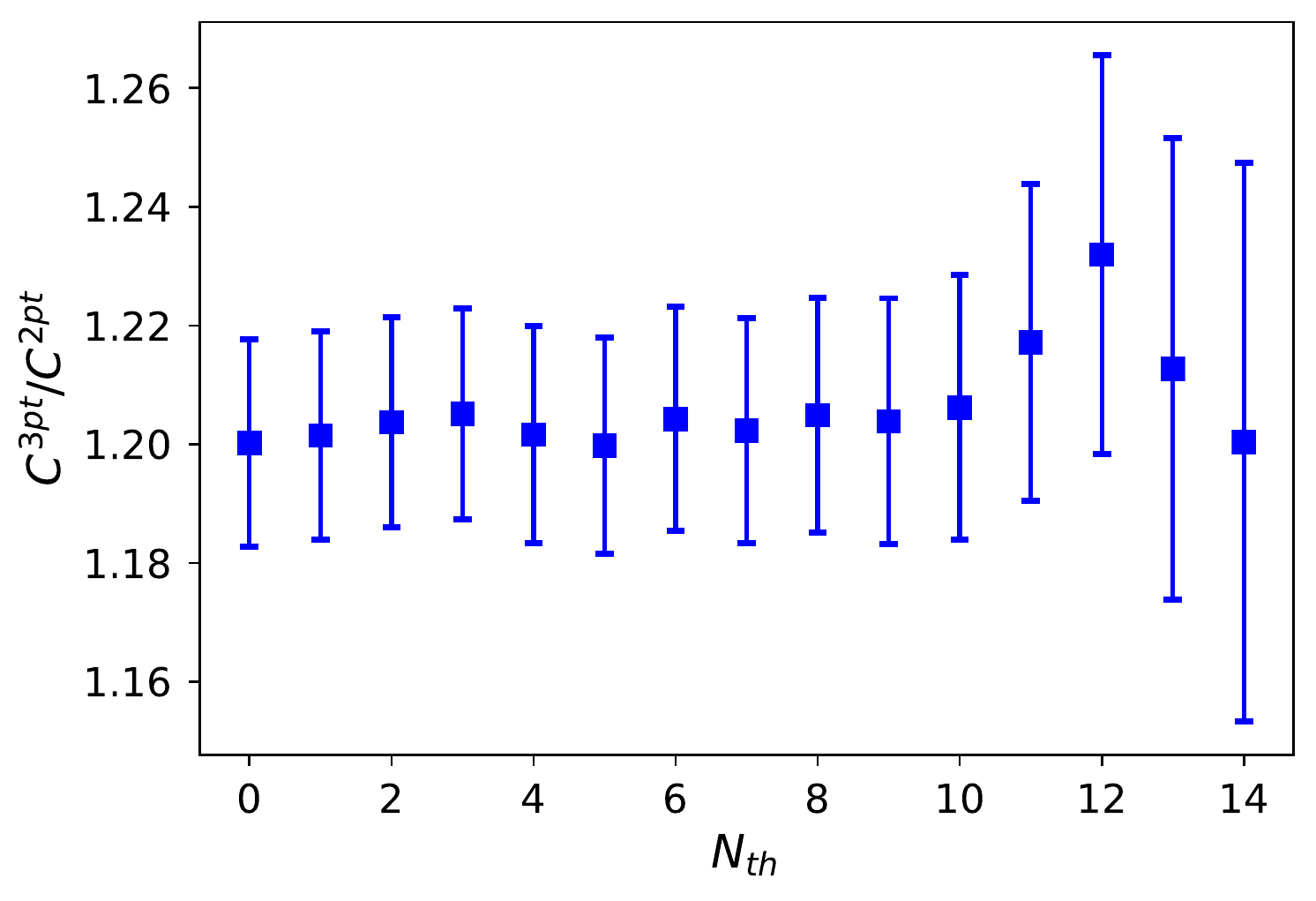}
 	\includegraphics[width=0.45\textwidth]{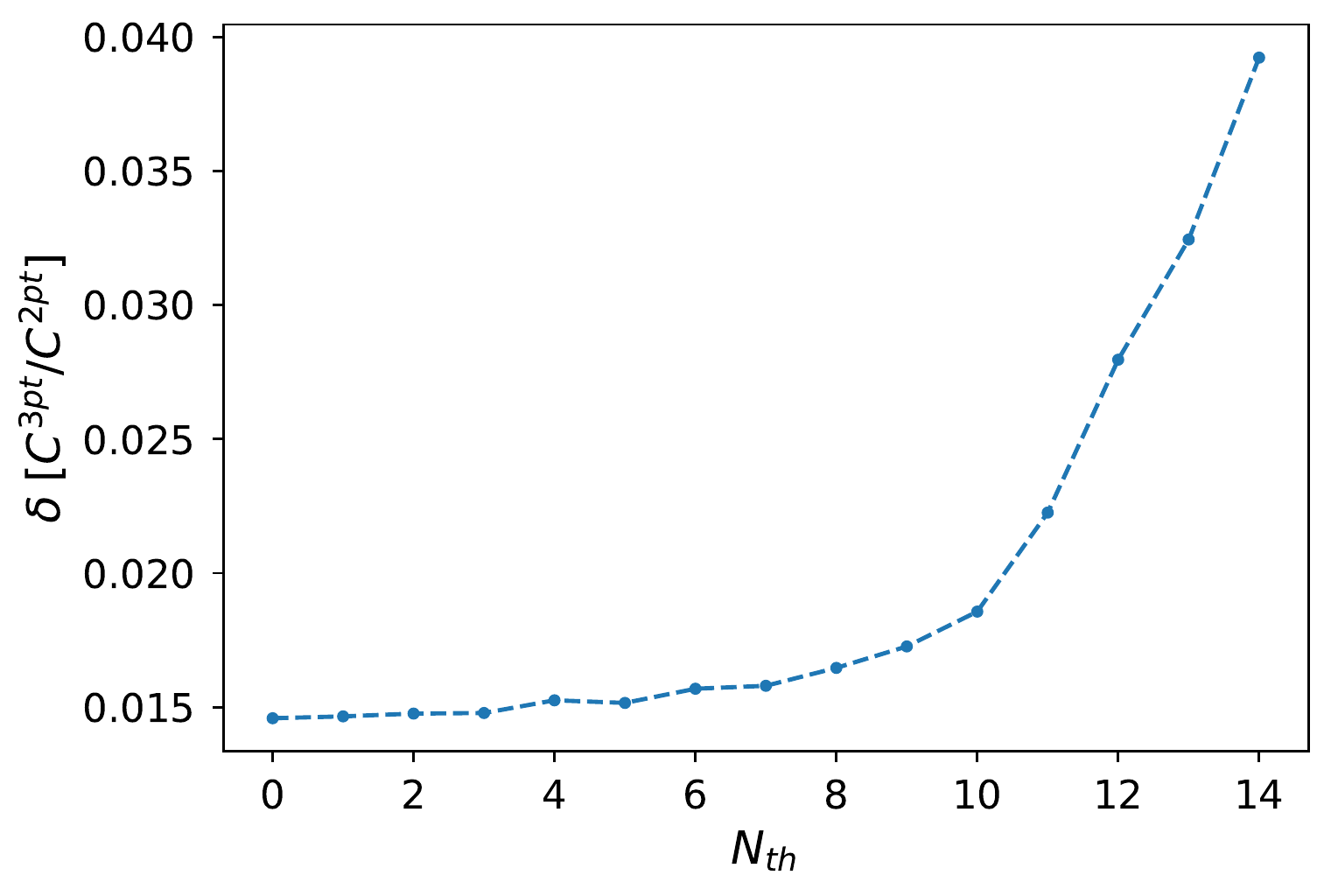}
 	\caption{
    The left panel shows the ratio between 3-point and 2-point functions,
$C^{\mathrm{3pt}}/C^{\mathrm{2pt}}$, as a function of $N_{\Lambda_s}$ (the
$x$-axis is labeled by $N_{\mathrm{th}}$). The 3-point functions are
     constructed using the sequential-source propagators. The time separation $t_i$ and $t_s$
     are fixed as $t_i=5$ and $t_s=10$. In the right panel, the uncertainty of
     the ratio as a function of $N_{\mathrm{th}}$ is shown.}
 	\label{fig:sequential}
 \end{figure}

In order to compare the field-sparsening results with the full-size ones, we use
the sequential-source propagators, which start from the source
$(t_0,\vec{x}_0)$, go through the current insertion point $(t_0+t_i,\vec{x})$ and
end at the
sink location $(t_0+t_s,\vec{x}_s)$.
We use 24 time slices for $t_0$. For each $t_0$, one random source is used and
the time $t_i$ and $t_s$ are fixed as
$t_i=5$ and $t_s=10$. In total we generate
24 sequential-source propagators for each configuration.
These propagators allow us to obtain the 3-point functions with 
$N_{\Lambda_0}=1$, $N_{\Lambda_t}=24$, 
$N_{\Lambda}=L^3$ and arbitrary values of $N_{\Lambda_s}$. By comparing the
lattice results at different $N_{\Lambda_s}$, one can estimate how large
the correlation among the sink points in the 3-point function. 
In the current study, we fixed $t_i$ and $t_s$.
The excited-state
effects will be taken into consideration later. In Fig.~\ref{fig:sequential}, it shows the
ratio between 3-point and 2-point functions,
$C^{\mathrm{3pt}}/C^{\mathrm{2pt}}$, as a function of $N_{\Lambda_s}$ (the
$x$-axis is labeled by $N_{\mathrm{th}}$).
We find that at $N_\mathrm{th}=10$ ($N_{\Lambda_s}=16$), the uncertainty of the
ratio $C^{\mathrm{3pt}}/C^{\mathrm{2pt}}$ only increases by 12\% compared to the
case of $N_\mathrm{th}=0$ ($N_{\Lambda_s}=L^3$). This is very consistent with
the observation in the effective masses of the 2-point functions.
Thus we can conclude that the field-sparsening method seems to work equally well for both
2-point and 3-point functions.

\begin{figure}
		\includegraphics[width=0.4\textwidth]{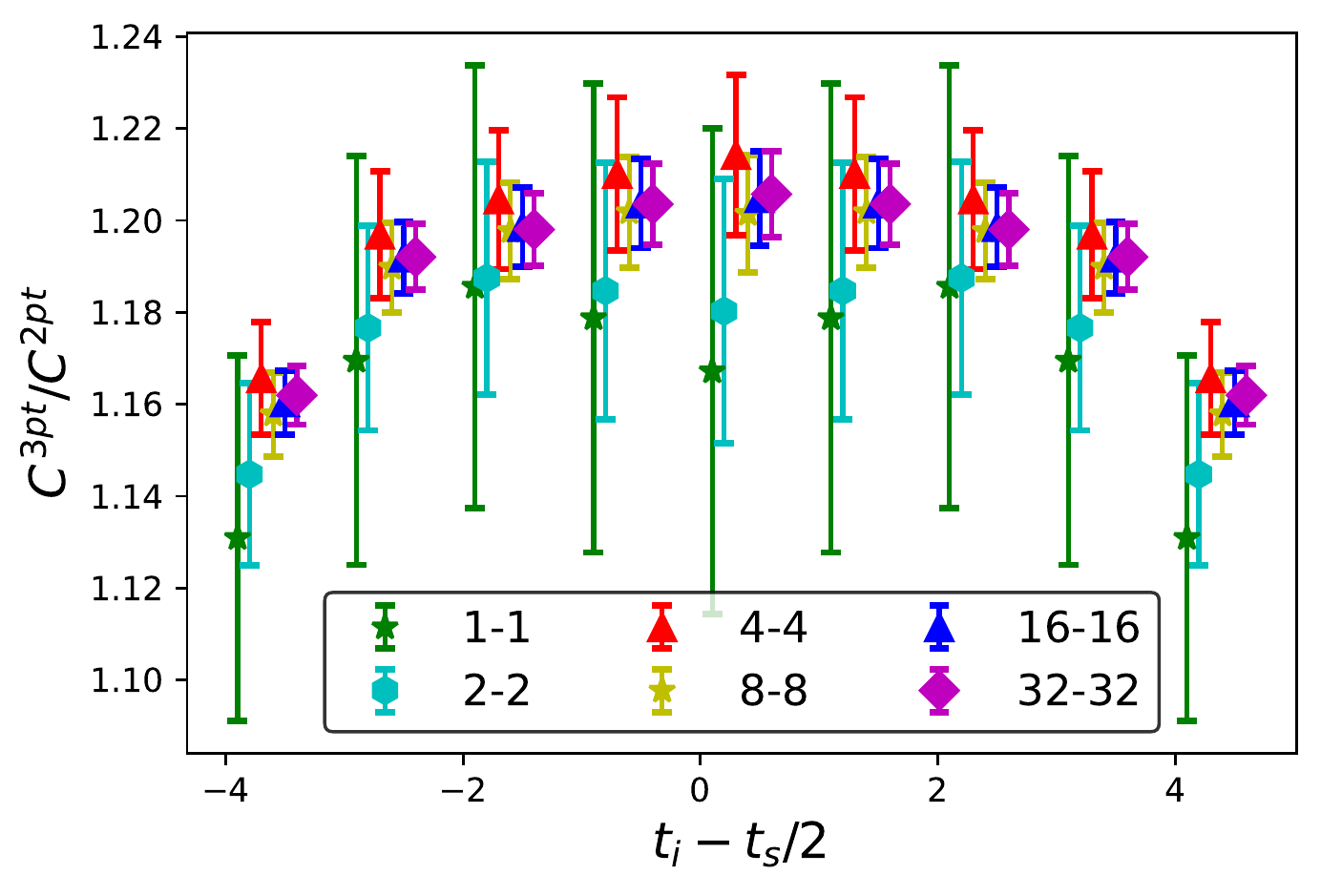}
		\includegraphics[width=0.4\textwidth]{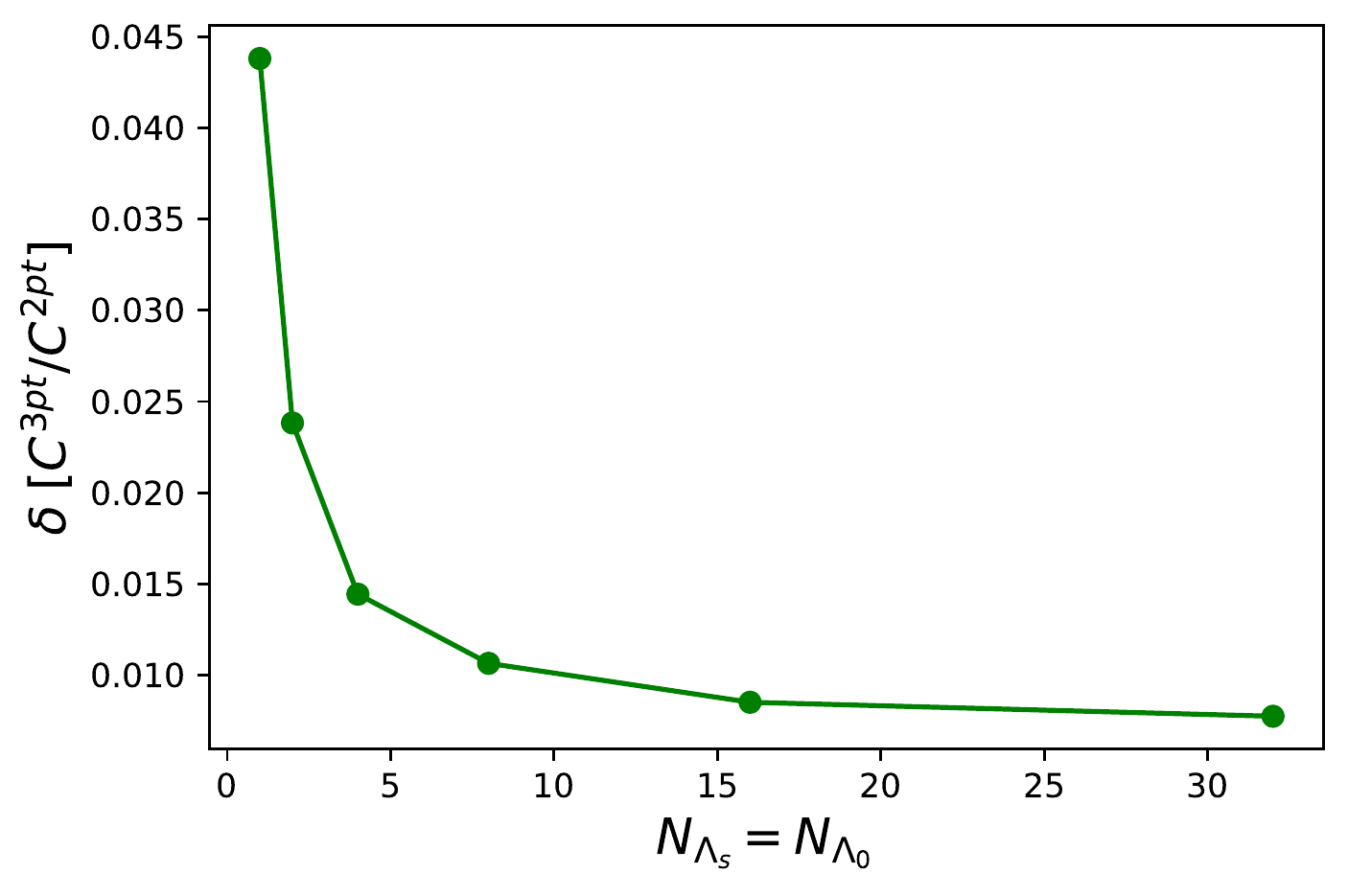}
	\caption{
        The left panel shows the results of $C^{\mathrm{3pt}}/C^{\mathrm{2pt}}$
as a function of $t_i-t_s/2$, where $t_s$ is chosen to be 10. The right panel
    shows the uncertainty of the ratio (with $t_s=10$ and $t_i-t_s/2=0$) as
   a function of $N_{\Lambda_0}=N_{\Lambda_s}$.
   In these two plots, we set $N_\Lambda = L^3$ .}
	\label{fig:compare_axial}
\end{figure}

As a next step, we want to determine the optimal values of $N_{\Lambda_0}$ and
$N_{\Lambda_s}$ for the source and sink location points.
Here the 3-point functions are constructed using the Gaussian-smeared propagators
only, which start from both source and sink
locations and end at the current insertion point
$(t_0+t_i,\vec{x})$. These Gaussian-smeared propagators are placed at 24 time
slices, which results in $N_{\Lambda_t}=24$. At each time slice we calculate
32 Gaussian-smeared propagators. It allows us to build the correlators with the
$N_{\Lambda_0}$-$N_{\Lambda_s}$ pair changing from $1$-$1$ to $32$-$32$.
We perform the summation of current insertion location $\vec{x}$ over the
whole spatial volume and have $N_{\Lambda}=L^3$. 
In the left panel of Fig.~\ref{fig:compare_axial}, we show the results of $C^{\mathrm{3pt}}/C^{\mathrm{2pt}}$
as a function of $t_i-t_s/2$, where $t_s$ is chosen to be 10 and the values of
$t_i-t_s/2$ vary in the range of $[-4,4]$. 
The data points with various $N_{\Lambda_0}$-$N_{\Lambda_s}$ pairs are plotted
using the different symbols.
In the right panel of Fig.~\ref{fig:compare_axial},
the statistical uncertainty of $C^{\mathrm{3pt}}/C^{\mathrm{2pt}}$ as a function of
$N_{\Lambda_0}=N_{\Lambda_s}$ are shown. 
We find that with $N_{\Lambda_0}$ changing from $1$ to $4$, the statistical
error drops relatively fast. From $N_{\Lambda_0}=4$ to $8$, the error decreases
slower. Due to the high correlation, the change of the uncertainty is very mild
from $N_{\Lambda_0}=8$ to 32. It is unnecessary to move on to larger
$N_{\Lambda_0}$ as one can expect that the precision at $N_{\Lambda_0}=N_{\Lambda_s}=8$ is close to the best precision using
the all-to-all setup ($N_{\Lambda_0}=N_{\Lambda_s}=L^3$). In practise, we
can choose $N_{\Lambda_0}=4$ or $8$ and invest the additional computational
resources in accumulating data from more gauge configurations.

 \begin{figure}[h]
 	\centering
 	\includegraphics[width=0.45\textwidth]{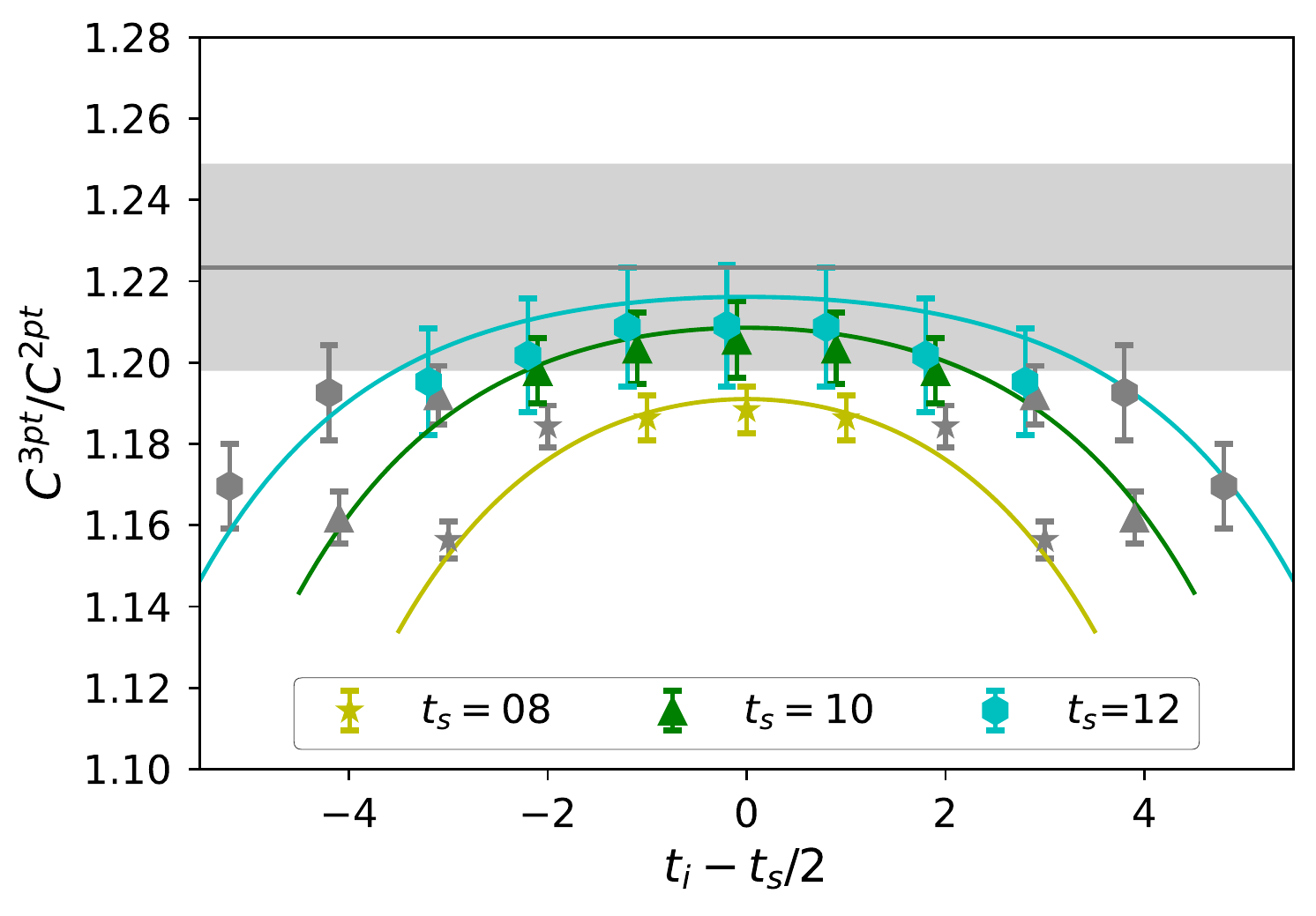}
     \caption{The ratio of $C^{\mathrm{3pt}}/C^{\mathrm{2pt}}$ as a function of
     $t_i-t_s/2$ together with a two-state fit. The shadowed band indicates the
     result of $g_A$ obtained from the fit.
     We use $N_{\Lambda_s} = N_{\Lambda_0} = 32$, $N_\Lambda=L^3$, and $N_{\Lambda_t} = 24$.
     }
    \label{fig:gA}
 \end{figure}

In Fig.~\ref{fig:compare_axial}, the ratio of
$C^{\mathrm{3pt}}/C^{\mathrm{2pt}}$ at $t_s=10$ is shown.
As a next step we add the lattice results at another two values of $t_s$ ($t_s=8$ and 12)
and vary $t_i-t_s/2$ in a range of $[-t_s+1,t_s-1]$. 
The corresponding
results are shown
in Fig.~\ref{fig:gA}. At large time separation, 
the time dependence of $C^{\mathrm{3pt}}/C^{\mathrm{2pt}}$ can be approximated
by a two-state form
as
\ba
\label{eq:excited_state}
\frac{C^{\mathrm{3pt}}(t_i,t_s)}{C^{\mathrm{2pt}}(t_s)}
=\frac{g_A+c_1 e^{-\Delta t_s}+c_2 (e^{-\Delta(t_s-t_i)}+e^{-\Delta
t_i})}{1+c_0e^{-\Delta t_s}},
\ea
where $\Delta$ is the energy difference between the excited state
and the ground state. The coefficients of $c_0$, $c_1$ and $c_2$ arise from
the excited-state contamination. 
We determine the values of $\Delta$ and $c_0$ from the
2-point function and then perform a two-state fit of
$C^{\mathrm{3pt}}/C^{\mathrm{2pt}}$ to Eq.~(\ref{eq:excited_state}) using three free parameters
$g_A$, $c_1$ and $c_2$. The lattice data for three $t_s$ are used in the fit
simultaneously with $t_i-t_s/2$ ranging from $[-t_s/2+3,t_s/2-3]$. As a final
result we
obtain the axial charge $g_A=1.223(25)$ at the pion mass $m_\pi\approx 350$ MeV
using 91 configurations and
$N_{\Lambda_s} = N_{\Lambda_0} = 32$, $N_\Lambda=L^3$, and $N_{\Lambda_t} = 24$
for each configuration.

  \begin{figure}[ht]
 	\centering
 	\includegraphics[width=0.45\textwidth]{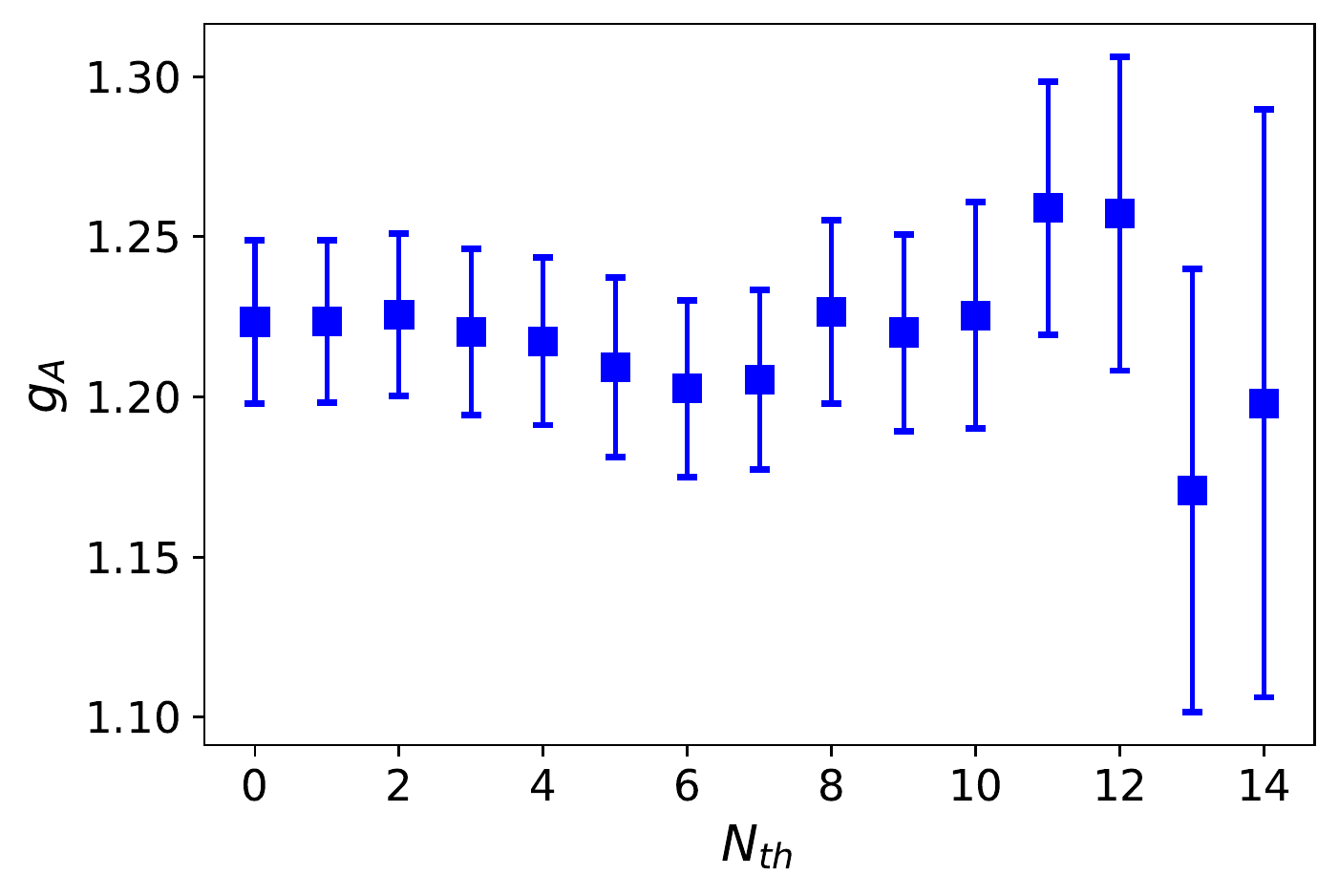}
 	\caption{The proton axial charge $g_A$ as a function of $N_{\mathrm{th}}$.
      Here $N_{\mathrm{th}}$ is related to the number of field points at the location of current
      insertion.
     We use $N_{\Lambda_s} = N_{\Lambda_0} = 32$ and $N_{\Lambda_t} = 24$.
      }
 	\label{fig:gA_summation}
 \end{figure}

As a last step, 
we calculate $g_A$ with different value of $N_{\Lambda}$, which is the number of
the field points at the location of current insertion $(t_0+t_i,\vec{x}_i)$.
In Fig.~\ref{fig:gA_summation}
we show $g_A$ as a function of $N_{\mathrm{th}}$.
By increasing $N_{\mathrm{th}}$ from 0 to 5 (or equavilently reducing
$N_{\Lambda}$ from $L^3$ to 108), 
we find that the uncertainty of $g_A$ only increases by 10\%.
Therefore, by using the field sparsening, one can reduce the size of the
Gaussian-smeared source point-sink propagators by a factor of 128
at the expense of 10\% increase in the statistical error of the correlation
function.

\section{Conclusion}

In this work we make an exploratory study on the field-sparsening methods. The
observables under investigation include the pion and proton 2-point
correlation functions and the proton axial charge $g_A$ involving the 3-point
functions. For the sparse-grid method, the results are not affected by the noise 
from field sparsening but
receive additional excited-state contamination from the higher-momenta states.
For the random field selection method, the situation is just the opposite. There is no
enhancement of the excited-state contamination, but the correlation functions
are affected by the noise from random selection. 
Fortunately, we confirm, in both numerical lattice results and a model
analysis mimicking pion correlator, that the noise from the random selection can be safely neglected.

In the calculation, we construct the correlation function using both
Gaussian-smeared operator and the point-source operator. We find that
Gaussian-smeared correlators do have higher correlation than the point-source ones.
At the expense of a $\sim$15\% increase in the statistical error, we can reduce
the number of field points by a factor of $\sim$100 for the point-source
operator and a factor of $\sim$1000 for the Gaussian-smeared operator. 
This is a surprisingly significant reduction in the size of the propagator. 
It also makes the storage of propagators much easier and saves the time for both I/O and 
quark contraction. Another interesting observation is that by using the field
sparsening methods, one can approach the precision of the all-to-all correlators 
with the modest cost of the computational resources.
Due to its high efficiency, we can foresee a vast application prospect of 
the field-sparsening methods proposed in this paper.

\begin{acknowledgments}

We thank ETM Collaboration for sharing the gauge configurations with us.
X.F. and L.C.J. gratefully acknowledge many helpful discussions with our colleagues from the
RBC-UKQCD Collaboration. 
X.F. and S.C.X. were supported in part by NSFC of China under Grant No. 11775002.
L.C.J acknowledges support by DOE grant DE-SC0010339.
Y.L. and C.L. are supported in part by CAS Interdisciplinary Innovation Team, NSFC of China under Grant No. 11935017, 
and the DFG and the NSFC through
    funds provided to the Sino-Germen CRC 110 ``Symmetries and the Emergence of
    Structure in QCD'', DFG grant no. TRR 110 and NSFC grant No. 11621131001.
    The calculation was carried out on TianHe-3 (prototype) at Chinese National Supercomputer Center in Tianjin.
\end{acknowledgments}

 \bibliography{field_selection}

\begin{thebibliography}{1}

\bibitem{Blum:2012uh}
T.~Blum, T.~Izubuchi, and E.~Shintani,
\newblock Phys. Rev. D {\bf 88}, 094503 (2013), 1208.4349.

\bibitem{Blum:2015gfa}
T.~Blum {\em et~al.},
\newblock Phys. Rev. D {\bf 93}, 014503 (2016), 1510.07100.

\bibitem{Foley:2005ac}
J.~Foley {\em et~al.},
\newblock Comput. Phys. Commun. {\bf 172}, 145 (2005), hep-lat/0505023.

\bibitem{Detmold:2019fbk}
W.~Detmold {\em et~al.},
\newblock (2019), 1908.07050.

\bibitem{Li:2010pw}
xQCD, A.~Li {\em et~al.},
\newblock Phys. Rev. D {\bf 82}, 114501 (2010), 1005.5424.

\bibitem{Martinelli:1988rr}
G.~Martinelli and C.~T. Sachrajda,
\newblock Nucl. Phys. B {\bf 316}, 355 (1989).

\bibitem{Alexandrou:2018egz}
C.~Alexandrou {\em et~al.},
\newblock Phys. Rev. D {\bf 98}, 054518 (2018), 1807.00495.

\bibitem{Aoki:2019cca}
Flavour Lattice Averaging Group, S.~Aoki {\em et~al.},
\newblock Eur. Phys. J. C {\bf 80}, 113 (2020), 1902.08191.

\end{thebibliography}
 
 \bibliographystyle{h-physrev}

\end{document}